\documentclass[12pt]{article}

\usepackage{amsmath}
\usepackage{amssymb}

\begin{document}

\newcommand{\id}{\relax{\rm 1\kern-.28em 1}}

\newcommand{\R}{\mathbb{R}}
\newcommand{\C}{\mathbb{C}}
\newcommand{\Z}{\mathbb{Z}}
\newcommand{\Hb}{\mathbb{H}}

\newcommand{\rSO}{\mathrm{SO}}

\newcommand{\cH}{\mathcal{H}}
\newcommand{\cL}{\mathcal{L}}

\newcommand{\fso}{\mathfrak{so}}
\newcommand{\fk}{\mathfrak{k}}
\newcommand{\fp}{\mathfrak{p}}

\newcommand{\e}{\epsilon}

\rightline{CERN-PH-TH/2006-036} \rightline{IFIC/06-12}

\vskip 3.5cm

\begin{center}{\LARGE \bf  Observations on the
Darboux coordinates\\ \medskip for rigid special
geometry}\end{center}

\vskip 1.5cm

\centerline{Sergio Ferrara$^{\flat}$ and \'Oscar
Maci\'a$^{\sharp}$}
 \vskip
1.5cm

\centerline{\it $^\flat$
 Physics Department, Theory Unit, CERN,}
\centerline{\small \it CH-1211 Geneva 23, Switzerland }
\centerline{{\footnotesize e-mail: Sergio.Ferrara@cern.ch}}
\medskip
\centerline{\it $^\flat$ INFN, Laboratori Nazionali di Frascati,}
\centerline{\small\it Via Enrico Fermi 40, I-00044 Frascati,
Italy}

 \bigskip

 \centerline{\it $^\sharp$
 Departament de F\'{\i}sica Te\`orica,
Universitat de Val\`encia and IFIC}
 \centerline{\small\it C/Dr.
Moliner, 50, E-46100 Burjassot (Val\`encia), Spain.}
 \centerline{{\footnotesize e-mail: Oscar.Macia@ific.uv.es}}


\vskip 1cm

\begin{abstract}
We exploit some relations which exist when (rigid) special
geometry is  formulated  in real symplectic special coordinates
$P^I=(p^\Lambda,q_\Lambda)\;,\; I=1,...,2n$. The central role of
the real $2n\times 2n$ matrix $M(\Re \mathcal{F},\Im \mathcal{F
})$, where $\mathcal{F}=\partial_\Lambda\partial_\Sigma F$ and $F$
is the holomorphic prepotential, is elucidated in the real
formalism. The property $M\Omega M=\Omega$, where $\Omega$ is the
invariant symplectic form, is used to prove several identities in
the Darboux formulation. In this setting the matrix $M$ coincides
with the (negative of the) Hessian matrix $H(S)=\frac{\partial^2
S}{\partial P^I\partial P^J}$ of a certain hamiltonian real
function $S(P)$, which also provides the metric of the special
K\"ahler manifold. When $S(P)=S(U+\bar U)$ is regarded as a
``K\"ahler potential''  of a complex manifold with coordinates
$U^I=\frac12(P^I+iZ^I)$,  it provides a K\"ahler metric of a
hyperk\"ahler manifold, which describes the
 hypermultiplet geometry obtained by $c$-map from the
original $n$-dimensional special K\"ahler structure.

\end{abstract}

 \vfill\eject

\section{Introduction}
Special geometry \cite{WP},\cite{WLP},\cite{S} plays an important
role in the description of the moduli space of  Calabi--Yau
compactifications for supergravity effective actions down to $D=4$
dimensions \cite{FS2} and also for more general compactifications
when fluxes \cite{PS},\cite{TV} of
different nature are turned on.\\
More interestingly,  special geometry has served as a basis to study
the so-called ``attractor mechanism''
\cite{FKS},\cite{FK},\cite{S2},\cite{M2} for black hole backgrounds
preserving at most four  supercharges.\\
In the rigid case
\cite{ST},\cite{G},\cite{seiberg},\cite{castellani}, special
geometry was an important tool for the Seiberg--Witten \cite{SW}
analysis of non-perturbative properties of $N=2$ super Yang--Mills
theories. An important ingredient in special geometry is the
existence of a flat symplectic bundle with structure group
$Sp(2n,\mathbb{R})\;,\;(Sp(2n+2,\mathbb{R})$ in the local case)
\cite{S},\cite{F},\cite{H},\cite{M2} over the special K\"ahler
manifold of complex dimension $n$, $n$ being the number
of vector multiplets in the theory.\\

\noindent It is the aim of the present work to elucidate some
properties of such a rich structure when Darboux real special
symplectic sections are adopted
\cite{CFG},\cite{F},\cite{H},\cite{M} for the description of
underlying mathematical structure. This description is particularly
suitable when background charges are introduced which are related to
fluxes of vector field-strength 2-forms over the space-time
manifold. Such a description has recently been used to simplify the
entropy area formula \cite{LWKM} for extremal black holes and its
relation to superstring theory \cite{OSV}. We will limit ourselves
to giving general results for the case of rigid special geometry but
many of these results can be extended to the local case which will
be described elsewhere.

\noindent We consider (rigid) special geometry in real special
(Darboux) coordinates $P^I=(p^\Lambda, q_\Lambda)\;,\;
\Lambda=1,...,n$ with the K\"ahler 2-form
\begin{equation}
\omega=i dq_\Lambda\wedge dp^\Lambda=\frac{i}{2}dP\wedge \Omega dP
\end{equation}
where $$\Omega=\left(
\begin{array}{cc}
0& -\mathbf{1}\\
\mathbf{1}&0
\end{array}\right)$$ is the symplectic invariant form.\\
The special geometry data turn out to be encoded in a real
``Hamiltonian" function $S(p,q)$, originaly introduced by Cecotti
et al. \cite{CFG} and by Freed \cite{F},  which is the Legendre
transform of the imaginary part of the holomophic prepotential
$F(X^\Lambda)$ of special geometry \cite{WLP},\cite{S},
\cite{ABCAFFM},\cite{CAF},\cite{CAFP}. The holomorphic symplectic
sections
\begin{equation}
V=(X^\Lambda,F_\Lambda)
\end{equation}
are related to the real variables as follows:
\begin{eqnarray}
\Re X^\Lambda=p^\Lambda \quad \Im
X^\Lambda=\phi^\Lambda=\frac{\partial S(p,q)}{\partial
q_\Lambda}\\
\Re F_\Lambda=q_\Lambda \quad \Im
F_\Lambda=\psi_\Lambda=-\frac{\partial S(p,q)}{\partial p^\Lambda}
\end{eqnarray}
If we encode the imaginary parts in the symplectic real vector
$I^I=(\phi^\Lambda,\psi_\Lambda)$, it turns out that
\begin{equation}
I^I=\Omega'^{IJ}\frac{\partial S}{\partial P^J}
\end{equation}
with $$\Omega'^{IJ}=-\Omega_{IJ}=\left(\begin{array}{cc}
0&\mathbf{1}\\-\mathbf{1}&0
\end{array}\right)$$ In this note we will show the special
property played by the $2n\times 2n$ real symmetric
positive-definite matrix \cite{CAF},\cite{FK}
\begin{equation}\label{m}
M(\Re \mathcal{F},\Im \mathcal{F})=\left(
\begin{array}{cc}
\Im \mathcal{F} + \Re \mathcal{F}  \Im \mathcal{F}^{-1}  \Re \mathcal{F} & - \Re \mathcal{F}  \Im \mathcal{F}^{-1}\\
-\Im \mathcal{F}^{-1} \Re \mathcal{F} & \Im \mathcal{F}^{-1}
\end{array}
\right)
\end{equation}
which in the real formulation turns out to be related to the
Hessian matrix
\begin{equation}\label{hessian} H_{IJ}(S)=\frac{\partial^2 S}{\partial P^I \partial P^J}
\end{equation}namely \begin{equation}\label{mh}
M(p,q)= -H(S)
\end{equation}
Note that $M$ is positive as a consequence of the fact that $\Im F$
is positive \cite{CAF}, \cite{FK} which is required by the
positivity of the metric (see Eq. (\ref{referee})). The matrix $M$
is known to play a special role when a background (symplectic)
charge vector $Q=(m^\Lambda, e_\Lambda)$ is introduced and a
``central charge" holomorphic function
\begin{equation}
Z=\langle Q,V\rangle=Q^T\Omega V=X^\Lambda e_\Lambda - m^\Lambda
F_\Lambda
\end{equation}
is defined.\\
In rigid special geometry\footnote{See ref. \cite{CAF2} for an
intrinsic definition.} the following identity holds (for the local
case, see later):
\begin{equation}
Q+i\Omega MQ= -2i g^{\Lambda \bar \Sigma}\partial_\Lambda V \bar
\partial_{ \Sigma}\bar Z
\end{equation}
Multiplying by $Q^T\Omega$
 on the left we get the ``central charge" potential function
\begin{equation}
\frac12 Q^TMQ= g^{\Lambda \bar \Sigma}\partial_\Lambda Z
\bar\partial_\Sigma \bar Z
\end{equation} where $g^{\Lambda \bar \Sigma}$ is the inverse of the
metric defined below.\\
\noindent The Hessian (\ref{hessian}) gives
also the metric of the original K\"ahler manifold
\begin{equation}\label{gk}
g_{\Lambda\bar \Sigma}dz^\Lambda \otimes d\bar
z^{\Sigma}=\frac{\partial^2 K}{\partial X^\Lambda
\partial \bar X^{\Sigma}}dX^\Lambda \otimes d\bar X^{\Sigma}
\end{equation}where special coordinates $z^\Lambda= X^\Lambda$ have been adopted and $$K=
-i(\bar X^\Lambda F_\Lambda-X^\Lambda \bar F_\Lambda)=-i\langle
V,\bar V\rangle$$ namely \begin{equation}\label{gh} g_{\Lambda\bar
\Sigma}dz^\Lambda \otimes d\bar z^{\Sigma}=-2 H_{IJ}dP^I\otimes
dP^J
\end{equation}
Another interesting observation is related to the $c$-map
hypermultiplet geometry as defined in references \cite{CFG},\cite{FS}.\\
By adopting real symplectic coordinates $Z^I=(\zeta^\Lambda,\tilde
\zeta_\Lambda)$ for the (other half) hypermultiplet coordinates, the
hyperk\"ahler metric has the form \cite{CFG},\cite{FS},\cite{AFTV}
\begin{equation}\label{sps}
g_{\Lambda\bar\Sigma}dz^\Lambda\otimes d\bar z^\Sigma+ 2
M_{IJ}(z,\bar z) dZ^I\otimes dZ^J
\end{equation}
Adopting Darboux coordinates for the K\"ahler manifold
$\mathcal{M}$, and because of (\ref{mh}) and (\ref{gh}), we then
have
\begin{equation}
-2\frac{\partial^2 S}{\partial P^I \partial P^J}(dP^I\otimes
dP^J+dZ^I\otimes dZ^J )
\end{equation}
By complexifying the Darboux coordinates as $U^I=\frac12 (P^I+ i
Z^I)$ we see that (\ref{sps}) is a K\"ahler metric with K\"ahler
potential \cite{H} \begin{equation}\label{ku}K(U,\bar U)=-8
S(U+\bar U )\end{equation} Note that, as expected from the results
of \cite{FS}, the metric (\ref{sps}) has $2n$ real isometries
\begin{equation} U^I\longmapsto U^I+i a^I
\end{equation}
In the ``complex" formulation of Cecotti et al. \cite{CFG}, where
the (second half of the) hypermultiplet coordinates were denoted
by $W_\Lambda$, the same  isometries were acting as
\begin{eqnarray}
W_\Lambda & \longmapsto & W_\Lambda + i \alpha_\Lambda\\
W_\Lambda & \longmapsto & W_\Lambda + (\Im
F)_{\Lambda\Sigma}\beta^\Sigma
\end{eqnarray}
and the K\"ahler potential was \cite{CFG}
\begin{equation}
N(X^\Lambda,W_\Lambda)=K(X,\bar X)+(\Im
\mathcal{F}^{-1})^{\Lambda\Sigma}(W_\Lambda+\bar
W_\Lambda)(W_\Sigma+\bar W_\Sigma)
\end{equation}
In the local case a particular choice of real coordinates has
recently  been  used \cite{LWKM},\cite{M} in conjunction with the
modification of the black hole entropy formula due to higher
curvature corrections and also in relating the entropy area
formula with the topological
string partition function \cite{OSV}.\\

The paper is organized as follows. In section 2 we give an
explicit derivation of the special K\"ahler metric in the Darboux
coordinates \cite{F},\cite{H} and, by using the properties of the
$M$ matrix we arrive at equations (\ref{hessian}), (\ref{gh}),
(\ref{gk}). In section 3 we discuss real special coordinates in
connection with the central charge function. We finally  comment
on some central charge relations and some metric differential
identities for  local special geometry.

\section{Complex and real special coordinates}
\subsection{ Rigid real sections and the functional $S(p,q)$}

Let $(X^\Lambda,F_\Lambda)$ be the rigid holomorphic symplectic
sections depending on the holomorphic coordinates $z^\Lambda$
\cite{CAFP}

\begin{equation}X^\Lambda=X^\Lambda(z)\qquad F_\Lambda=F_\Lambda(z)\end{equation}

\noindent Under some general  assumptions we can take $X^\Lambda$ as
the holomorphic coordinates $X^\Lambda=z^\Lambda$ such that
\begin{equation}F_\Lambda(X)=\frac{\partial F(X)}
{\partial X^\Lambda}\end{equation} for a suitable  function $F(X)$.\\

\noindent The holomorphicity condition reads
\begin{equation}\label{holo}\bar
\partial_{\Sigma}X^\Lambda=0\qquad\qquad \bar\partial_{\Sigma}F_\Lambda=0\end{equation}

\noindent Decomposing $(X^\Lambda,F_\Lambda)$ in terms of real and
imaginary parts we can write the holomorphic symplectic sections in
terms of real variables
\begin{equation}X^\Lambda=p^\Lambda+i\phi^\Lambda \qquad F_\Lambda=
q_\Lambda+i\psi_\Lambda\end{equation}

\noindent Define the function $L= \Im F>0$, the imaginary part of
$F(X)$. Then it can be proved that $(q_\Lambda,\phi^\Lambda)$ and
$(p^\Lambda,\psi_\Lambda)$ are pairs of conjugate variables for $L$
\begin{equation}\label{invert}
q_\Lambda=\frac{\partial L}{\partial
\phi^\Lambda}\end{equation}\begin{equation}
\psi_\Lambda=\frac{\partial L}{\partial p^\Lambda}\end{equation}

\noindent We  perform a Legendre transform on $L$ of the form
\begin{equation}S(p,q)=q_\Lambda\cdot\phi^\Lambda(p,q)-L(p,\phi(p,q))\end{equation}
where we have to invert equation (\ref{invert}) to write
$\phi=\phi(p,q)$. Then, the next set of equations follows:
\begin{eqnarray} \phi^\Lambda&=&\frac{\partial S
}{\partial q_\Lambda}\\- \psi_\Lambda&=&\frac{\partial S}{\partial
p^\Lambda}\end{eqnarray} Our change of coordinates
$(p^\Lambda,\phi^\Lambda)\mapsto (p^\Sigma,q_\Sigma)$ is of the form
\begin{eqnarray}\label{coord1}p^\Sigma =\delta^\Sigma_\Lambda p^\Lambda\qquad
q_\Sigma =q_\Sigma(p^\Lambda,\phi^\Lambda)
\end{eqnarray}
and, accordingly, the inverse change of coordinates
$(p^\Sigma,q_\Sigma)\mapsto (p^\Lambda,\phi^\Lambda)$ is
\begin{eqnarray}\label{coord2}
p^\Lambda=\delta^\Lambda_\Sigma p^\Sigma \qquad \phi^\Lambda
=\phi^\Lambda(p^\Sigma,q_\Sigma)
\end{eqnarray}
\noindent  In the following we use the notation $$\frac{\partial
f(x,y)}{\partial x}\Big|_y=\frac{\partial f}{\partial x} \quad
\qquad \frac{\partial f(x,y)}{\partial y}\Big|_x=\frac{\partial
f}{\partial y}$$  where the role of $f(x,y)$ will be played by
$q_\Sigma(p,\phi),\; \phi^\Lambda(p,q)$ as defined in Eqs.
(\ref{coord1}), (\ref{coord2}).\\

\noindent Let $\mathbb{J}$ be the Jacobian matrix. It follows from
$\mathbb{J}(\mathbb{J})^{-1}=\mathbb{\textbf{1}}=(\mathbb{J})^{-1}\mathbb{J}$
that \begin{eqnarray}\label{ji} \frac{\partial q_\Sigma}{\partial
\phi^\Lambda }\frac{\partial \phi^\Lambda}{\partial
q_\Gamma}=\delta^\Sigma_\Gamma=\frac{\partial q_\Sigma}{\partial
\phi^\Lambda} \frac{\partial^2S}{\partial q_\Lambda\partial
q_\Gamma}
\end{eqnarray}
i.e.\begin{equation} \frac{\partial q_\Sigma}{\partial
\phi^\Lambda}=\left(\frac{\partial^2 S}{\partial q_\Lambda\partial
q_\Sigma}\right)^{-1}=\frac{\partial^2 L}{\partial \phi^\Sigma
\partial \phi^\Lambda}
\end{equation}and we also have
\begin{eqnarray}\label{36}
\frac{\partial q_\Sigma}{\partial
p^\Lambda}\delta^\Lambda_\Gamma+\frac{\partial q_\Sigma}{\partial
\phi^\Lambda}\frac{\partial \phi^\Lambda}{\partial
p^\Gamma}=\frac{\partial q_\Sigma}{\partial
p^\Lambda}\delta^\Lambda_\Gamma+\left(\frac{\partial^2 S}{\partial
q_\Sigma\partial q_\Lambda}\right)^{-1}\frac{\partial^2 S
}{\partial q_\Lambda \partial p^\Gamma}=0
\end{eqnarray}

\noindent In fact, equations (\ref{ji}) to (\ref{36}) follow not
only from the Jacobian but also from the holomorphicity conditions
(\ref{holo}). These also imply
\begin{eqnarray}\label{38}
\frac{\partial q_\Sigma}{\partial\phi^\Lambda}=-\frac{\partial
\psi_\Sigma}{\partial p^\Lambda}-\frac{\partial q_\Gamma}{\partial
p^\Lambda}\frac{\partial \psi_\Sigma}{\partial q_\Gamma }
\end{eqnarray} where $\psi=\psi(p,q)$ and its partial derivatives
are computed following the notation explained before. The
analyticity of $L=\Im F$, for $F$ holomorphic, also imply
$$\frac{\partial^2 L}{\partial \phi^\Lambda\partial\phi^\Sigma}+
\frac{\partial^2 L}{\partial p^\Lambda \partial p^\Sigma}=0\qquad
\qquad \frac{\partial^2 L}{\partial \phi^\Lambda
\partial p^\Sigma }= \frac{\partial^2 L}{\partial p^\Lambda \partial
\phi^\Sigma}$$ In terms of $S$ Eq. (\ref{38}) can be written as
\begin{equation}\label{jff}
\left(\frac{\partial^2 S}{\partial q_\Sigma\partial
q_\Lambda}\right)^{-1}=\frac{\partial^2 S}{\partial p^\Sigma
\partial p^\Lambda}+\frac{\partial q_\Gamma}{\partial
p^\Lambda}\frac{\partial^2 S}{\partial q_\Gamma \partial p^\Sigma }
\end{equation} These relations  will be useful when simplify the metric in
the next section.

\subsection{K\"ahler potential and metric}

In this section we show that the K\"ahler potential
 $K$ can be expressed directly in terms of $S(p,q)$. Then, the metric can
also be expressed directly in terms of $S(p,q)$, since
\begin{equation}
g=g_{\Lambda\bar\Sigma}dX^\Lambda\otimes d\bar X^\Sigma
\end{equation}
with
\begin{equation}
g_{\Lambda\bar\Sigma}=\frac{\partial^2 K}{\partial
X^\Lambda\partial \bar X^{\Sigma}}
\end{equation}

\noindent In the rigid case the K\"ahler potential is
\begin{equation}
K=i(X^\Lambda \bar F_\Lambda-\bar X^\Lambda F_\Lambda)
\end{equation} The K\"ahler form then is
\begin{equation}\omega=-\frac14 \partial_\Lambda
\bar
\partial_{\Sigma }K \; dX^\Lambda\wedge d\bar X^\Sigma= -\frac12
\Im F_{\Lambda \Sigma}\; dX^\Lambda\wedge d\bar X^\Sigma
\end{equation}
so that
\begin{equation}\omega= i dq_\Lambda\wedge dp^\Lambda \end{equation}
is a symplectic form. The metric will be
\begin{equation}\label{referee}g_{\Lambda \bar \Sigma}=2 \Im F_{\Lambda\Sigma}>0\end{equation}
Using equations (\ref{ji}) to (\ref{jff}), one arrives at
\begin{equation}
g_{\Lambda\bar \Sigma}=-2\left( \frac{\partial^2 S}{\partial
p^\Lambda \partial p^\Sigma}+\frac{\partial q_\Gamma}{\partial
p^\Sigma}\frac{\partial^2 S}{\partial q_\Gamma \partial
p^\Lambda}\right)=-2\left( \frac{\partial^2 S}{\partial
q_\Lambda\partial q_\Sigma}\right )^{-1}
\end{equation}
Note that  the symmetry properties  $F_{\Lambda
\Sigma}=F_{\Sigma\Lambda}$ imply
\begin{eqnarray}\label{sym1}
\frac{\partial q_\Lambda}{\partial p^\Sigma}&=&\frac{\partial
q_\Sigma}{\partial
p^\Lambda}\\
\frac{\partial q_\Gamma}{\partial \phi^\Lambda}\frac{\partial^2
S}{\partial q_\Gamma \partial p^\Sigma}&=&\frac{\partial
q_\Gamma}{\partial \phi^\Sigma}\frac{\partial^2 S}{\partial
q_\Gamma\partial p^\Lambda}\\\label{sym2} \frac{\partial
q_\Gamma}{\partial p^\Lambda}\frac{\partial^2 S}{\partial q_\Gamma
\partial p^\Sigma}&=&\frac{\partial q_\Gamma}{\partial
p^\Sigma}\frac{\partial^2 S}{\partial q_\Gamma
\partial p^\Lambda}
\end{eqnarray}
which will help in the simplifications.\\

\noindent The change of variables in the differentials gives
\begin{eqnarray}
dX^\Lambda&\otimes& d\bar X^\Sigma =
\left(\delta^\Lambda_\Gamma\delta^\Sigma_\Delta+\frac{\partial^2 S
}{\partial q_\Lambda\partial p^\Gamma}\frac{\partial^2 S
}{\partial q_\Sigma\partial p^\Delta}+i\left(\frac{\partial^2 S
}{\partial q_\Lambda\partial
p^\Gamma}\delta^\Sigma_\Delta-\frac{\partial^2 S }{\partial
q_\Sigma\partial p^\Delta}\delta^\Lambda_\Gamma
\right)\right)dp^\Gamma\otimes dp^\Delta \nonumber \\&+& \left(
\frac{\partial^2 S }{\partial q_\Lambda\partial
q_\Delta}\frac{\partial^2 S }{\partial q_\Sigma\partial
p_\Gamma}+\frac{\partial^2 S }{\partial q_\Lambda
\partial p_\Gamma}\frac{\partial^2 S }{\partial q_\Sigma \partial
q_\Delta}+i\left(\frac{\partial^2 S }{\partial q_\Lambda \partial
q_\Delta}\delta^\Sigma_\Gamma-\frac{\partial^2 S }{\partial
q_\Sigma\partial q_\Delta}\delta^\Lambda_\Gamma\right)
 \right)dp^\Gamma\otimes
dq_\Delta\nonumber
\\&+&\left(\frac{\partial^2 S }{\partial q_\Lambda \partial
q_\Gamma}\frac{\partial^2 S }{\partial q_\Sigma \partial
q_\Delta}\right)dq_\Gamma \otimes dq_\Delta
\end{eqnarray}

\noindent Finally, using again equations (\ref{ji}) to (\ref{jff}),
and (\ref{sym1}) to (\ref{sym2}) the following expression is
obtained:
\begin{equation}\label{oscar} g(p,q)=-2 \left( \frac{\partial^2
S}{\partial p^\Lambda \partial p^\Sigma} dp^\Lambda\otimes
dp^\Sigma + 2 \frac{\partial^2 S}{\partial p^\Lambda \partial
q_\Sigma} dp^\Lambda\otimes dq_\Sigma + \frac{\partial^2
S}{\partial q_\Lambda\partial q_\Sigma}dq_\Lambda \otimes
dq_\Sigma\right)
\end{equation}

\noindent Equation (\ref{oscar}) implies that the Hessian matrix
(\ref{hessian}) is negative-definite.\\ Comparison of (\ref{oscar})
with a different evaluation of the metric in Darboux coordinates
(\ref{referee2}) in section 2.4 , will allow us to prove that
$M=-H$, as asserted in the introduction.

\subsection{The K\"ahler form}
In rigid special geometry the symplectic holomorphic vector
$V=(X^\Lambda, F_\Lambda)$ defines the K\"ahler form through the
formula \cite{F}
\begin{equation}
\Omega_{IJ} dV^I\wedge d\bar V^{J}=-dX^\Lambda\wedge d\bar
F_\Lambda+dF_\Lambda\wedge d\bar X^\Lambda
\end{equation}
by writing $X^\Lambda=p^\Lambda+i\phi^\Lambda\;,\;
F_\Lambda=q_\Lambda+i\psi_\Lambda$, we find
\begin{equation}
-2dp^\Lambda\wedge dq_\Lambda-2 d\phi^\Lambda\wedge d\psi_\Lambda
\end{equation}
On the other hand, if we compute (using the property
$dX^\Lambda\wedge dF_\Lambda=0$)
\begin{eqnarray}
d(X^\Lambda+\bar X^\Lambda)\wedge d(F_\Lambda+\bar
F_\Lambda)=dX^\Lambda \wedge d\bar F_\Lambda+d\bar X^\Lambda\wedge
dF_\Lambda\nonumber
\\=(dX^\Lambda\wedge d\bar F_\Lambda-dF_\Lambda\wedge
d\bar X^\Lambda)= 4dp^\Lambda\wedge dq_\Lambda
\end{eqnarray}

\noindent the following relation must hold,
\begin{equation}\label{dRedIm}
dp^\Lambda\wedge dq_\Lambda = d\phi^\Lambda\wedge d\psi_\Lambda
\end{equation}
We postpone in proving (\ref{dRedIm}) but simply observe that
\begin{equation}\label{reomega}
\omega=\frac{i}{4} \Omega_{IJ}dV^I\wedge d\bar
V^J=\frac{i}{2}\Omega_{IJ}dP^I\wedge dP^J=idq_\Lambda\wedge
dp^\Lambda
\end{equation}

\subsection{The K\"ahler metric}

We now consider the K\"ahler metric in complex coordinates. Let us
first note a basic identity satisfied by the complex symplectic
differential $$dV=(dX^\Lambda,
dF_\Lambda)=(dX^\Lambda,F_{\Lambda\Sigma}dX^\Sigma)$$ and the $M$
matrix defined in (\ref{m}). It is easy to see that
\begin{equation}\label{basic}
M dV = i \Omega dV
\end{equation}
and also to recall that $M$ is real, symmetric and symplectic,
i.e. that it satisfies \begin{equation} \label{basic2}M\Omega
M=\Omega
\end{equation}
The basic identity (\ref{basic}) allows us to prove that
\begin{eqnarray}\label{dvmfdv} d\bar V M(\mathcal{F})dV=i d\bar V \Omega dV = i \langle d\bar V,dV\rangle
= i(d\bar X^\Lambda, d\bar
F_\Lambda)\left(\begin{array}{c}-dF_\Lambda\nonumber \\dX^\Lambda
\end{array}\right)\\=i(dX^\Lambda d\bar F_\Lambda - d\bar
X^\Lambda dF_\Lambda)=i dX^\Lambda d\bar X^\Sigma (\bar F_{\Lambda
\Sigma}- F_{\Lambda\Sigma})=2 dX^\Lambda d\bar X^\Sigma \Im
F_{\Lambda\Sigma}
\end{eqnarray}

Therefore the (positive-definite) K\"ahler metric  is
\begin{eqnarray}
2\Im F_{\Lambda\Sigma}=\partial_\Lambda \bar\partial_{\Sigma} K
\qquad K=-i\langle V, \bar V\rangle
\end{eqnarray}
Let us now consider the Darboux (special) coordinates
$$P^I=(\Re X^\Lambda, \Re F_\Lambda)=(p^\Lambda, q_\Lambda)$$ and
$$I^I=(\Im X^\Lambda, \Im F_\Lambda)=(\phi^\Lambda,
\psi_\Lambda)$$ such that $$dV^I= dP^I + i dI^I \qquad d\bar V^I=
dP^I-i dI^I$$ In components we have
\begin{equation}
dV^Id\bar V^J=(dP^I+ i dI^I)(dP^J-i dI^J)= (dP^I dP^J +
dI^IdI^J)+i(dI^IdP^J-dP^IdI^J)
\end{equation}

\begin{equation}\label{mdvmdp}
M_{IJ}dV^I d\bar V^J = M_{IJ}(dP^I dP^J+ dI^IdI^J)
\end{equation}

\noindent We now compute $dI^I dI^J$ by using the following property
\begin{eqnarray}\label{omegavector}
I^I=\left( \begin{array}{c}\phi^\Lambda \\
\psi_\Lambda\end{array}\right)=\left( \begin{array}{c}\frac{\partial S}{\partial q_\Lambda} \\
-\frac{\partial S}{\partial
p^\Lambda}\end{array}\right)=\Omega'^{IJ}\frac{\partial
S}{\partial P^J}=-\Omega \frac{\partial S}{\partial P}
\end{eqnarray}
where $\Omega'^{IJ}=-\Omega_{IJ}$ so that $\Omega' \Omega =
\mathbf{1}$. It then follows that
\begin{equation}\label{spomegahp}
dI^I= \Omega ' H dP
\end{equation}
where $H$ is the  (real symmetric) Hessian of $S(P)$. By inserting
(\ref{spomegahp}) into (\ref{mdvmdp}) we obtain
\begin{equation}\label{referee2}
M_{IJ} dV^I \otimes d\bar V^J=(M-H\Omega M \Omega H)_{IJ}dP^I\otimes
dP^J
\end{equation}
by comparing Eq. (\ref{oscar}) with  (\ref{referee2}) we get
\begin{equation}\label{mdvdv}
M_{IJ}dV^I\otimes d\bar V^J= -2 H_{IJ}dP^I\otimes dP^J
\end{equation} Eqs. (\ref{referee2}) and (\ref{mdvdv}) then imply that $M=-H$.
The same argument can be used for the K\"ahler form to prove
equation (\ref{dRedIm}). In fact
\begin{equation}
\Omega_{IJ}dV^I\wedge d\bar V^J = \Omega_{IJ}(dP^I\wedge dP^J+
dI^I\wedge dI^J)
\end{equation}
by use of (\ref{omegavector}), (\ref{spomegahp}) we get
\begin{equation}
\Omega_{IJ}dV^I\wedge d\bar V^J=dP^I\wedge
dP^J(\Omega-M\Omega\Omega\Omega M)_{IJ}
\end{equation}
and, using $\Omega^3=-\Omega$, $M\Omega M=\Omega$, we get
\begin{equation}\label{dvomega}
\Omega_{IJ} dV^I\wedge d\bar V^J=2 \Omega_{IJ}dP^I\wedge dP^J
\end{equation}
which is nothing but equation (\ref{reomega})

\subsection{The hyperk\"ahler metric}

The Darboux coordinates give also a  striking simplification
\cite{H} of the metric of the real  $4n$-manifold of the
hypermultiplet geometry, which is obtained by the $c$-map
introduced by \cite{CFG} and which associates to any special
K\"ahler manifold of dimension $n$ an hyperk\"ahler manifold of
complex dimension $2n$. Furthermore, as discussed in \cite{FS},
this manifold has at least $2n$ isometries ($2n+3$ in the local
case which come from the $c$-map construction \cite{FS}). The
hypermultiplet geometry, as described in \cite{CFG}, has a metric
of the form\footnote{Note that because of property (\ref{basic2})
if we lower the indices $Z_I$ the second term in (\ref{gdzm})
becomes $2(M^{-1})^{IJ}dZ_I\otimes dZ_J$. This agrees with the
hypermultiplet metric as given in \cite{M}}
\cite{CFG},\cite{FS},\cite{AFTV}
\begin{equation}\label{gdzm}
g_{\Lambda\bar \Sigma}dz^\Lambda \otimes d\bar z^{\Sigma}+
2M_{IJ}(z,\bar z)dZ^I\otimes dZ^J
\end{equation}
where $Z^I=(\zeta^\Lambda,\bar \zeta_\Lambda)$ are $2n$ real
coordinates associated to a symplectic real vector $Z$ and
$g_{\Lambda \bar\Sigma}$ is the original K\"ahler metric.\\By
adopting Darboux coordinates and noticing that $M=-H$ equation
(\ref{dvomega}) takes the simple form
\begin{equation}\label{sps2}
-2 \frac{\partial^2 S}{\partial P^I \partial P^J}(dP^I\otimes
dP^J+d Z^I\otimes dZ^J)=\frac{\partial^2 K}{\partial U^I \partial
\bar U^J}dU^I\otimes d\bar U^J
\end{equation}
 It is immediate
to see that (\ref{sps2}) is a K\"ahler metric, for a complex
 $2n$-dimensional manifold with $2n$ complex coordinates given by
\begin{equation}
U^I=\frac12(P^I+ i Z^I)
\end{equation}
and K\"ahler potential given by (\ref{ku}). Interestingly enough
the hamiltonian function $S(P)$ has a double role. In the original
symplectic manifold $\mathcal{M}$ of complex dimension $n$, its
Hessian matrix in Darboux coordinates is the metric on the
manifold. Considered as a function of ``complex coordinates"
$U^I$, it is the K\"ahler potential on the cotangent bundle (of
the special manifold $\mathcal{M}$)
with real dimension $4n$.\\

\noindent Note the obvious $2n$ isometries \begin{equation}
U^I\longmapsto U^I+ i a^I
\end{equation}
as implied by the analysis of \cite{FS}.

\section{Central charges and special coordinate identities in complex and real coordinates}
In the present section we discuss identities and relations of
special geometry in presence of a background charge real symplectic
vector $Q=(m^\Lambda, e_\Lambda)$. In terms of the special geometry
complex sections the ``central charge'' function is given by
\begin{equation} Z=\langle Q, V\rangle = Q^T\Omega V
\end{equation}
where scalar symplectic products are understood. In local special
geometry $Z$ is  a section of a $U(1)$ bundle over $\mathcal{M}$
and it is conveniently written in terms of symplectic sections
over $U(1)$ \cite{CAFP},\cite{CAF},\cite{FK}
$$Z^L=\langle Q,V^L\rangle$$
\medskip
\begin{equation}
V^L=(L^\Lambda, M_\Lambda) \qquad i\langle V^L,\bar V^L\rangle= 1
\end{equation}
so that \begin{equation} K= - log(i(\bar X^\Lambda F_\Lambda -
X^\Lambda \bar F_\Lambda))
\end{equation}
In the rigid case the K\"ahler potential is  rather  given by
\begin{equation}
K=-i\langle V,\bar V\rangle = -i (\bar X^\Lambda F_\Lambda -
X^\Lambda \bar F_\Lambda)
\end{equation}
so that $$K_{\Lambda \bar \Sigma } =  2 \Im \emph{F}_{\Lambda
\Sigma}>0$$ Note that in the local case $\Im
\mathcal{F}_{\Lambda\Sigma}$ is a matrix of lorentzian signature
\cite{CAF} with $n$ positive and one negative eigenvalues, while
the matrix $\mathcal{N}$
$$\mathcal{N}_{\Lambda\Sigma}=\bar h_{I\Lambda}(\bar f^{-1})^I_\Sigma$$
(where $\bar h_{I\Lambda},\;\bar f_{I\Lambda}$ are $(n+1)\times
(n+1)$ matrices which are the components of the sections $\bar
D_{i} \bar V,\; I=1,...,n$, and $V,\; I=0$) is negative-definite
(in the rigid case $\mathcal{N}\mapsto  \mathcal{\bar F}$ so that
$\Im \mathcal{N}\mapsto - \Im\mathcal{F}$, and $\Im \mathcal{F}$
becomes positive-definite and is
an $n\times n$ instead of an $(n+1)\times(n+1)$ matrix).\\

In the local case, if we define the $U(1)$ covariant derivatives
$D_i Z^L$ ($\bar D_{\bar i}Z^L=0$), $D_i V^L$ in terms of the
$U(1)$ sections $Z^L$ and $V^L$, then the following identity is
true\footnote{Taking the real part of (\ref{qomegam}) we obtain
the identity used in \cite{BFC},\cite{FK1},\cite{FK2}. This
identity has recently been generalized in the presence of more
general fluxes in Ref. \cite{D}.} \cite{BFM}:
\begin{equation}\label{qomegam}
Q-i\Omega M(\mathcal{N}) Q = -2i \bar V^L Z^L-2i g^{i\bar j}D_i
V^L \bar D_{\bar j}\bar Z^L
\end{equation}

\noindent with the (local) special geometry identity
\begin{equation}\label{mvomegav}
M(\mathcal{N})V^L= i\Omega V^L
\end{equation}
where $M(\mathcal{N})$  is the same matrix as in (\ref{m})
but with $\mathcal{F}\longmapsto \mathcal{N}$.\\
Note that if  in (\ref{qomegam}) we take the scalar product with
$Q$, since $\langle Q, Q\rangle=0$ we obtain the black-hole
potential as \cite{FK}
\begin{equation}\label{bh} -\frac12 Q^T M(\mathcal{N})Q= |Z^L|^2+ D_i Z^L \bar
D_{\bar j} \bar Z^L  g^{i\bar j}= V_{BH}
\end{equation}
On the other hand, by multiplying by $V$ and using the fact that
$MV=i \Omega V$, we obtain $$\langle Q, V\rangle = Z$$ The rigid
formula that replaces (\ref{qomegam}) is
\begin{equation}\label{qomegam2}
Q+ i \Omega M(\mathcal{F}) Q = -2i g^{i\bar j}\partial_i V \bar
\partial_{\bar j}\bar Z
\end{equation}\label{qtmq}
which implies the rigid formula \begin{equation}\label{VQTMQ}
V=\frac12 Q^T M(\mathcal{F})Q= g^{i\bar j} \partial_i Z \bar
\partial_{\bar j}\bar Z
\end{equation}
Note that (\ref{VQTMQ}), with respect to (\ref{bh}), loses the
graviphoton charge contribution and it is identical to the
$N=1$ rigid formula for chiral multiplets of superpotential $Z$.\\
This formula coincides with the superpotential contribution to the
$N=2$ potential considered in Ref. \cite{APT}. From
(\ref{qomegam2}) we also see that at a supersymmetric extremum
$\partial_i Z=0$
implies $Q=0$, something which is different from the local case.\\
The local supersymmetric attractor point $$D_i Z^L=0$$ gives,
instead, the so called ``BPS attractor equations'':
\begin{eqnarray}
Q=-i (\bar V^L Z^L - V^L \bar Z^L)\\
\Omega M(\mathcal{N}) Q = \bar V^L Z^L+ V^L \bar Z^L
\end{eqnarray}
The rigid identities (\ref{qomegam2}) can be  written for real
Darboux symplectic special coordinates  by noticing that
\begin{equation}
Z=\langle Q,V\rangle=p^\Lambda e_\Lambda -m^\Lambda q_\Lambda + i
(\phi^\Lambda-m^\Lambda \psi_\Lambda)=Z(Q,P,I)
\end{equation}
By using now the property that $$I=\Omega'\frac{\partial
S}{\partial P}$$ we get \begin{equation} \frac{\partial
Z}{\partial P^I}=-\Omega Q + i H Q = \tau Q
\end{equation}
with $\tau= -\Omega+ i H$, $\tau=-\tau^\dag$.\\
Note that, as a consequence of the fact  that $H\Omega H= \Omega$
we have  $$\tau \Omega \tau = 2 \tau \qquad \tau \Omega \tau^T=0$$
The potential becomes \begin{equation} V(P,Q)=-\frac12 Q^THQ
\end{equation}

\subsection{Local special geometry}
We finally discuss further identities of local special geometry.
Since
\begin{equation}
V=(L^\Lambda, M_\Lambda=\mathcal {N}_{\Lambda \Sigma }L^\Sigma)
\end{equation} and also
\begin{equation}M_\Lambda=F_{\Lambda\Sigma}L^\Sigma\end{equation} (whenever
$\emph{F}_\Lambda=\partial_\Lambda \emph{F}$) it is also true
that, in addition to (\ref{mvomegav}), we have
\begin{equation}
M(\mathcal{F})V^L=i\Omega V^L
\end{equation} by further use of the identity
\begin{equation}
dM_\Lambda=\emph{F}_{\Lambda\Sigma\Delta}X^\Sigma
dL^\Delta+\emph{F}_{\Lambda\Sigma}dL^\Sigma=F_{\Lambda\Sigma}dL^\Sigma
\end{equation}
(because $F_{\Lambda\Sigma\Delta}X^\Sigma=0$) it is also true, as
in the rigid case, that
\begin{equation}M(\mathcal{F})dV^L = i \Omega dV^L\end{equation}  By using the
definition \begin{equation}D_iV^L=\partial V^L + \frac12
K_iV^L\end{equation} we have
\begin{equation}M(\mathcal{F}) D_i V^L = i\Omega D_i V^L \qquad (\bar
D_{\bar i} V^L=0)\end{equation} where $D_i$ is a $U(1)$
covariant derivative.\\
The K\"ahler metric on the manifold is given by \cite{CAFP}
\begin{equation}g^L=i\langle\bar D \bar V^L, D V^L\rangle= i \bar
D\bar V^L \Omega D V^L\end{equation} By using the previous
relations we get the equivalent expression
\begin{equation}\label{gl}g^L=( \bar D \bar V^L
M(\mathcal{F})DV^L)
\end{equation}
Formula (\ref{gl}) is the local analogue  of (\ref{dvmfdv}) and it
will be useful  to formulate local special geometry in Darboux
real special coordinates.

\section{Acknowledgement}
We would like to thank M.A. Lled\'o for enlightening
discussions.\\
S.F. would like to thank the Theoretical Physics Department at the
University of Valencia where part of this work was performed. S.F.
was partially supported by funds of the INFN-CICYT bilateral
agreement. The work of S.F. has also been supported in part by the
European Community Human Potential Program under contract
MRTN-CT-2004-005104 ``Constituents, fundamental forces and
symmetries of the Universe'', in association with INFN Frascati
National
Laboratories  and by the D.O.E. grant DE-FG03-91ER40662, Task C.\\
O.M. wants to thank the Department of Physics, Theory Division at
CERN for its kind hospitality during the realization of this work.
The work of O.M. has been supported by an FPI fellowship from the
Spanish Ministerio de Educaci\'on y Ciencia (M.E.C.), through the
grant FIS2005-02761 (M.E.C.) and EU FEDER funds, by the
Generalitat Valenciana, contracts GV04B-226, GV05/102 and by the
EU network MRTN-CT-2004-005104 ``Constituents, fundamental forces
and symmetries of the Universe''.

\end{document}